\documentclass[twocolumn,prb,showpacs,preprintnumbers,amsmath,amssymb]{revtex4-1}
\usepackage{graphicx}
\usepackage{epstopdf}
\usepackage{dcolumn,color}
\usepackage{bm}
\begin{document}

\title{Valence fluctuations and empty-state resonance for Fe adatom on a surface}
\author{S.N. Iskakov$^{1}$, V. V. Mazurenko$^{1}$,  M.V. Valentyuk$^{1,2}$ and A. I. Lichtenstein$^{1,2}$}
\affiliation{$^{1}$Theoretical Physics and Applied Mathematics Department, Ural Federal University, 620002 Ekaterinburg, Russia \\
$^{2}$ Hamburg University, Germany, Hamburg, Jungiusstra{\ss}e 9}

\date{\today}

\begin{abstract}
We report on the formation of the high-energy empty-state resonance in the electronic spectrum of the iron adatom on the Pt(111) surface.  By using the combination of the first-principles methods and the finite-temperature exact diagonalization approach, we show that the resonance is the result of the valence fluctuations between atomic configurations of the impurity. 
Our theoretical finding is fully confirmed by the results of the scanning tunneling microscopy measurements [M.F. Crommie et al., Phys. Rev. B 48, 2851 (1993)]. In contrast to the previous theoretical results obtained by using local spin density approximation, the paramagnetic state of the impurity in the experiment is naturally reproduced within our approach.
This opens a new way for interpretation of STM data collected earlier for metallic surface nanosystems with iron impurities. 
\end{abstract}

\maketitle

\section{Introduction} The modern scanning tunneling microscopy (STM) experiments \cite{Wiesendanger1} are aimed to reveal, identify and operate excitations of surface nanosystems, which is of crucial importance for understanding  the basic phenomena in quantum physics such as Kondo resonance,\cite{pruschke} magnetic anisotropy, exchange interactions\cite{otte} between adatoms and for constructing novel information devices on the atomic level. Typically, these excitations present the smallest energy scale in the system, they are of a few millivolts and can be reproduced by solving Heisenberg-type models.\cite{rudenko}

However, the picture of the STM experiment in the case of an adatom on a metallic surface is not complete without considering high-energy resonances of hundreds meV.\cite{Meier} They carry information concerning inter-orbital charge or spin excitations in the impurity \cite{Mazurenko,Mazurenko2} and can be used to study electronic structure of the adatom. 
In this sense there is one important example of such high-energy excitations that is a pronounced peak at 0.5 eV above the Fermi level in the tunneling spectra of the iron adatom deposited on Pt(111),\cite{Crommie}  Pd(111) \cite{Wasniowska} or W(110) \cite{Bode} surfaces.  Importantly, it was proposed that the peak can be used to identify iron species in surface nanosystems.

The theoretical description of the high-energy excitations such as 0.5 eV peak for iron adatom on a metallic surface is a complex methodological and numerical problem. Within the Lang's model \cite{Lang} that establishes the connection between experimental STM spectrum and the local density of states at the site of the adatom, this peak can be attributed to the $4s$ states of  the Fe adatom. However, the $ab-initio$ calculations based on the density functional theory do not confirm such a scenario. As we will show below the non-spin-polarized density of states exhibits a $3d$ peak in the iron adatom spectral function at the Fermi level (Fig.\ref{LDA}). Above E$_{F}$ the density of states is a decreasing function without any features. The intensity of the $4s$ states is very small. 

As it was demonstrated in Ref.\onlinecite{Wasniowska} the experimentally observed excitations can be reproduced in the framework of the spin-polarized local density approximation (LSDA) calculations. For that one should take into account the shift of the spin-down $3d$ states to higher energies above the Fermi level due to the spin splitting. Then within the Tersoff-Hamann approach\cite{Tersoff} the STS is associated with the scattering of the $s-p$ electrons at spin-down $d_{z^2}$ states.\cite{Blugel} 

Thus to describe the experiment one should define some ordered magnetic state of the system in the $ab-initio$ calculations. As we will show below, this scenario is fulfilled in the case of Fe/Pt(111).  At the same time, the experimental conditions (low temperatures and zero magnetic field) correspond to the paramagnetic state of Fe/Pt(111). To describe the system in this regime, a realistic five-orbital impurity Anderson model is constructed and solved by using the finite-temperature exact diagonalization  method. The correlated density of states shows a peak at 0.5 eV above the Fermi level in accordance with the experiment.\cite{Crommie} The composition analysis of the eigenvectors has shown that the 0.5 eV resonance is originated from the valence fluctuations, which are combined effect of the intra-atomic exchange coupling on the impurity and strong hybridization with the surface states. We also show that the position of the resonance is sensitive to the coupling with other adsorbates.  

\section{LDA and LSDA results} 
The first step of our investigation was to define an equilibrium atomic structure of the Fe/Pt(111) nanosystem. For that we performed first-principles molecular dynamics simulations by using the Vienna \textit{ab initio} simulation package (VASP) 
\cite{PhysRevB.47.558,PhysRevB.49.14251,Kresse199615,PhysRevB.54.11169} within local density approximation (LDA).
In these calculations the energy cutoff of 300 eV in the plane-wave basis
construction and the energy convergence criteria of $10^{-7}$ eV were used. The atomic positions were relaxed with residual
forces less than 0.01 eV/{\AA}. We used the PAW-PBE exchange-correlation potential as described in Ref.\onlinecite{PhysRevB.59.1758}. 

\begin{figure}[!t]
\centering
   \includegraphics[width=0.45\textwidth]{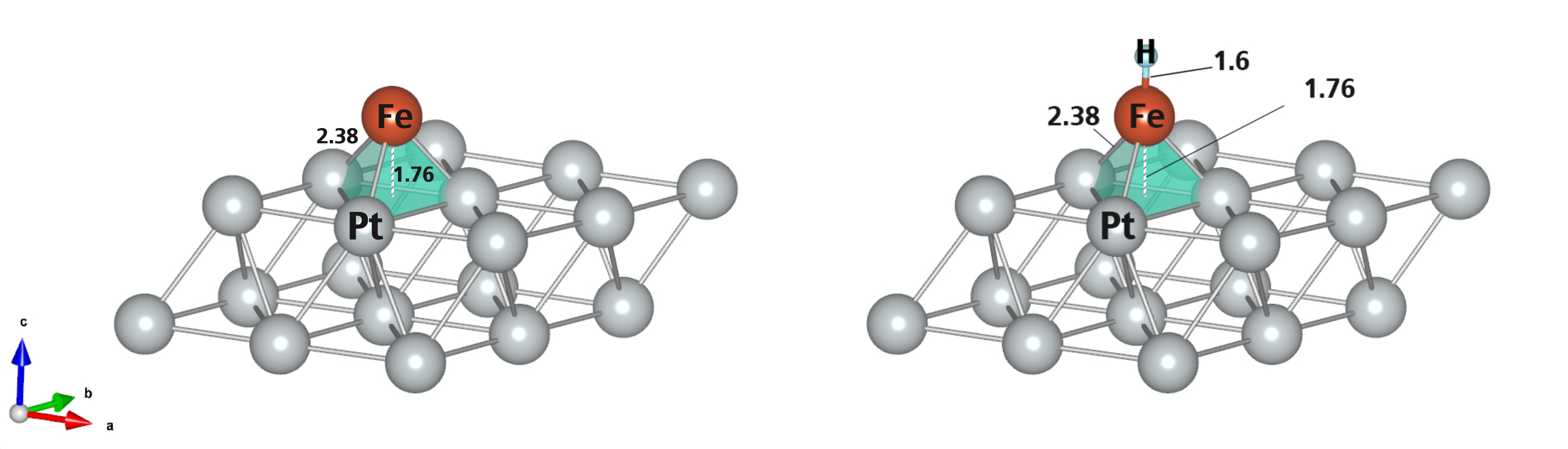}
 \caption{Schematic representation of the Fe/Pt(111) (left) and FeH/Pt(111) (right) nanosystems simulated in this work.}
\label{structure}
\end{figure}

The simulations of the atomic structure of Fe adatom on the Pt(111) surface were carried out  within a supercell approach.
The supercell contains a three-layered ($4\times4$) Pt(111) surface, iron atom and vacuum region of 13.4 \AA. A simplified illustration of the simulated Fe/Pt(111) system is given by Fig.\ref{structure} (left). The lattice constant for Pt lattice was chosen to be
3.92 {\AA} that is the experimental value of the lattice constant for the bulk fcc Pt. \cite{PhysRevB.77.184425}  The unit cell parameters were fixed during relaxation procedure. The obtained vertical distance between Fe atom and Pt surface of 1.61 {\AA} is in good agreement with the
reported values. \cite{MPIAnnualRep} 

\begin{figure}[!b]
\centering
   \includegraphics[width=0.49\textwidth]{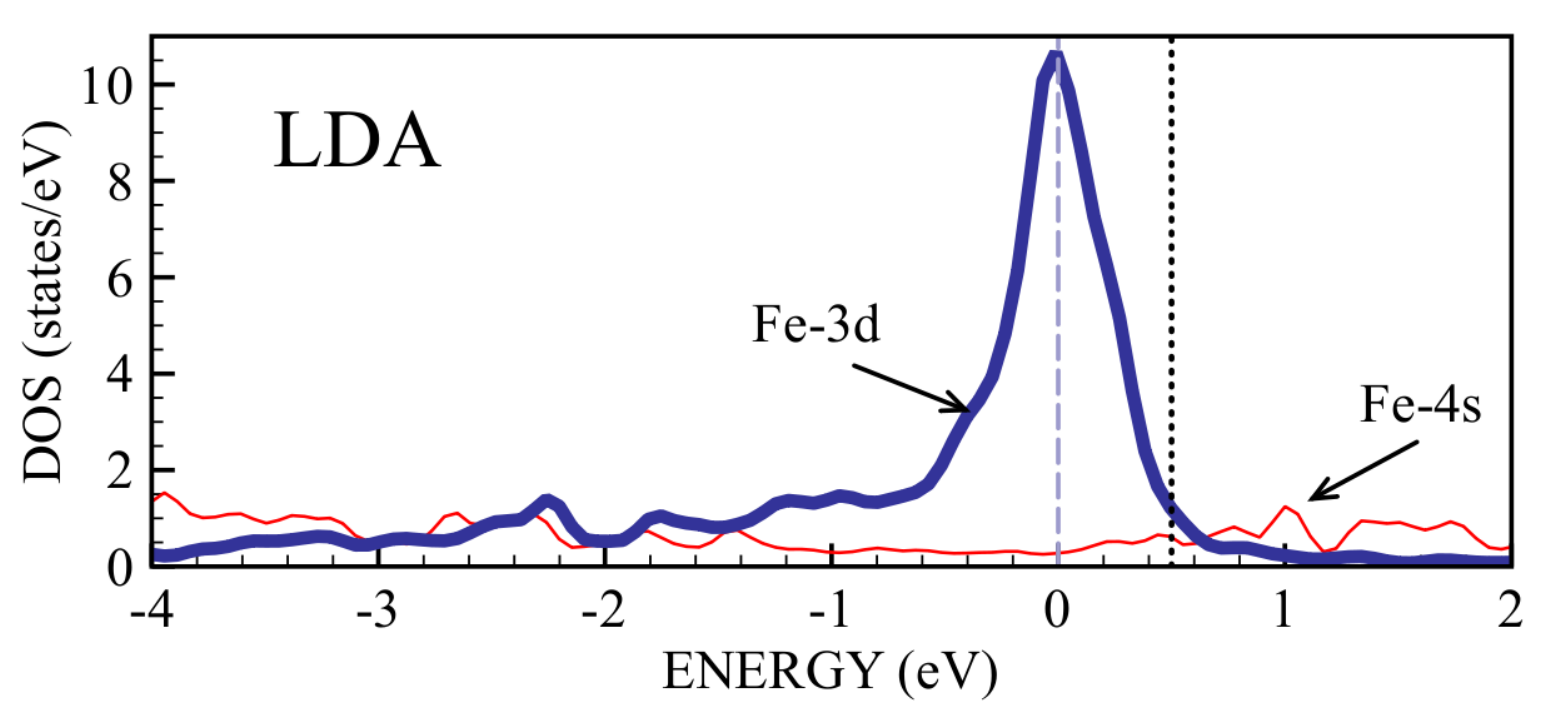}
 \caption{Partial densities of states obtained from LDA calculations. Blue bold and red thin lines denote $3d$ and $4s$ states of the iron adatom. The intensity of the Fe-4s is multiplied by ten. The dashed line corresponds to the Fermi level. The dotted line denotes the energy of the experimental resonance.}
\label{LDA}
\end{figure}

Fig.\ref{LDA} gives the partial densities of states obtained by using the local density approximation. There is a peak at the Fermi level. The width of the peak is about 1 eV that is much larger than one would expect for a Kondo system.
In contrast to the results obtained for Co/Pt(111) (Ref.\onlinecite{Mazurenko}) we do not observe a strong orbital polarization of the LDA spectra.
The integration of the density of states gives the Fe-3d occupation of 6.5.  One can see that LDA density of states does not reveal any feature at the energy of the experimental resonance of 0.5 eV. $4s$ states of iron can be also excluded from the consideration since they give a negligible contribution to the spectral function close to the Fermi level.  As we will show below the obtained LDA spectra can be used to extract the hopping integrals between impurity and surface states. The latter is important to construct a realistic Anderson model for Fe/Pt(111) system. 

\begin{figure}[!b]
\centering
   \includegraphics[width=0.49\textwidth]{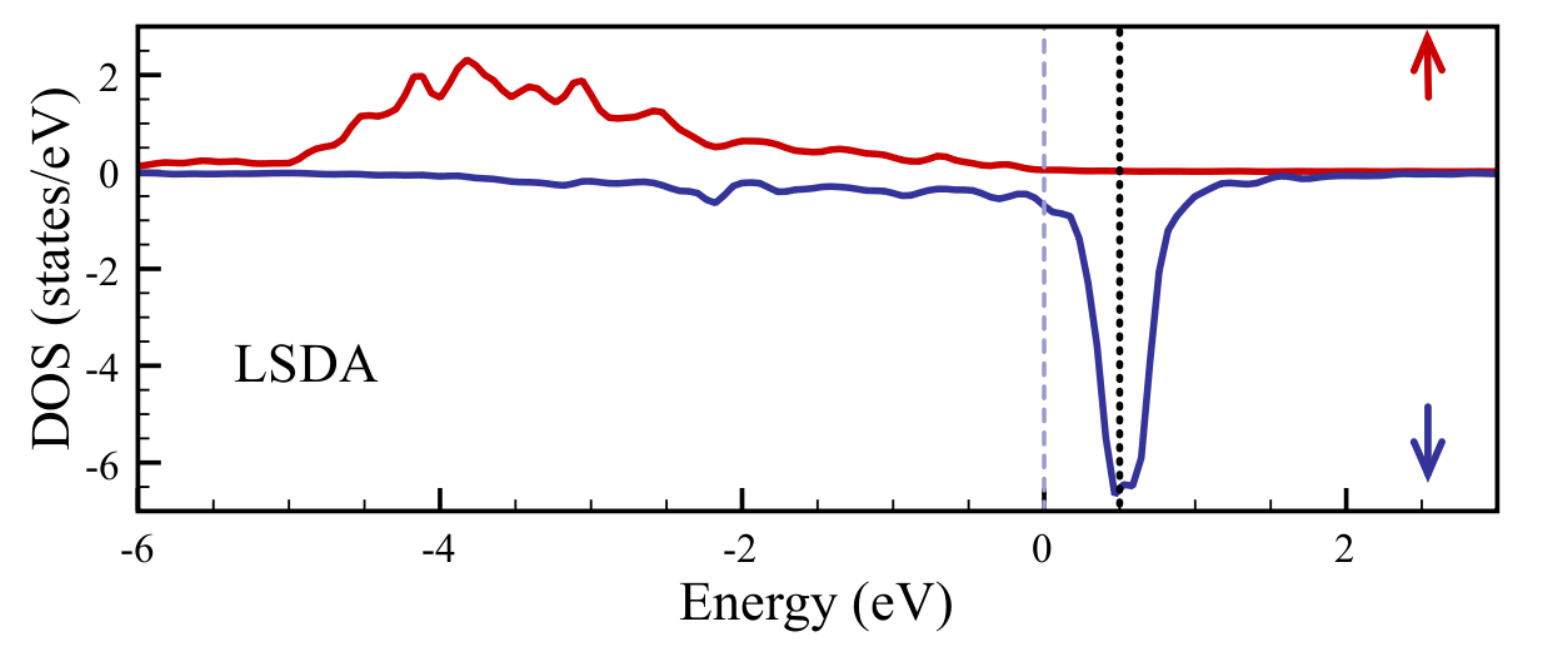}
 \caption{Spin-resolved densities of states of the iron adatom obtained from LSDA calculations. The dashed line corresponds to the Fermi level. The dotted line denotes the energy of the experimental resonance.}
\label{LSDA}
\end{figure}

The account of the spin-polarization within local spin density approximation leads to a significant change of the physical properties of Fe/Pt(111). In agreement with the results of previous works the spin splitting produces the peak at 0.5 eV in the spin-down channel (Fig.\ref{LSDA}). We obtain the magnetic solution with moments of the $3d$, $4s$ and $4p$ iron shells that are $M_{d}$ = 3.2 $\mu_{B}$, $M_{s}$ = 0.04 $\mu_{B}$ and $M_{p}$ = 0.034 $\mu_{B}$, respectively.
The value of the total moment, 3.3 $\mu_{B}$ is in good agreement with the previous results \cite{Blugel} and smaller than 4 $\mu_{B}$ one would expect for the isolated iron being in the $d^6$ atomic configuration.
A strong hybridization of the Fe and Pt states results in a partial magnetization of the surface, the induced magnetic moment of the surface can be estimated as 1.6 $\mu_{B}$. 
We observe the peak for spin-down $3d$ states of iron at the energy of the experimental resonance. These electronic and magnetic properties of Fe/Pt(111) agree with that reported in the previous works.\cite{Minar}

From Fig.\ref{LSDA} one can see that the gravity center of the occupied spin-up states shifts from the Fermi level to the energies of about -3.5 eV. They are strongly hybridized with the surface state and becomes completely delocalized in comparison with the LDA picture. At the same time, the width of the LSDA peak above the Fermi level is smaller then that in LDA, which indicates the increase of the localization of the spin-down $3d$ orbitals. The spin splitting that can be estimated from Fig.\ref{LSDA} is in reasonable agreement with model estimate $I_{d} (\langle n^{\uparrow}_{d} \rangle - \langle n^{\downarrow}_{d} \rangle)$, where $I_{d}$ is the Stoner parameter of about 1 eV. Interestingly, the number of the $3d$ electrons obtained in the LSDA solution, $\langle n^{LSDA}_{d} \rangle$ = 6.13 is smaller than the LDA value of 6.5. Since the spin-up states are almost occupied, the change of the total number of the electrons is related to the change of the hybridization of the iron spin-down states with the surface.

Thus the 0.5 eV resonance observed in the STM experiment \cite{Crommie} can be reproduced and explained by using the results of the LSDA calculations, however, for that one should assume the non-zero magnetization of the impurity.
In a real experiment \cite{Crommie} the impurity is in the paramagnetic state and below we present the results of Anderson model simulations that describe the Fe/Pt(111) system without magnetic ordering.  

\section{Anderson model} 
To take into account the paramagnetic state and dynamical electron-electron correlations in Fe/Pt(111) we have constructed and solved the following Anderson model 
\begin{eqnarray}
&\mathcal{H}_{Fe} = \sum_{i} (\epsilon_{i} - \mu )n_{i\sigma} +
\sum_{p\sigma}\epsilon_{p}c^{\dagger}_{p\sigma}c_{p\sigma} \nonumber \\ 
&+ \frac{1}{2} \sum_{ijkl\sigma\sigma'} U_{ijkl}d^{\dagger}_{i\sigma}d^{\dagger}_{j\sigma'}d_{l\sigma'}d_{k\sigma}
\nonumber \\
&+ \sum_{i p \sigma} \left( V_{ip}d^{\dagger}_{i\sigma}c_{p\sigma} + H.c. \right),
\end{eqnarray}
where $\epsilon_{i}$ and $\epsilon_{p}$ are energies of the impurity and surface states, $d^{\dagger}_{i\sigma}$ ($d_{i\sigma}$)
and $c^{\dagger}_{p\sigma}$($c_{p\sigma}$) are the creation (annihilation) operators for impurity and surface electrons, 
$V_{ip}$ is the hopping integral between impurity and surface states, $\mu$ is the chemical potential and $U_{ijkl}$ is the Coulomb matrix
element. The impurity orbital index $i$ ($j$, $k$, $l$) runs over the $3d$-states ($xy$, $yz$, $3z^2 - r^2$, $xz$, $x^2 - y^2$).
A realistic simulation of the Fe/Pt(111) system requires an accurate definition of these Hamiltonian parameters. 

\begin{figure}[!t]
\centering
   \includegraphics[width=0.34\textwidth]{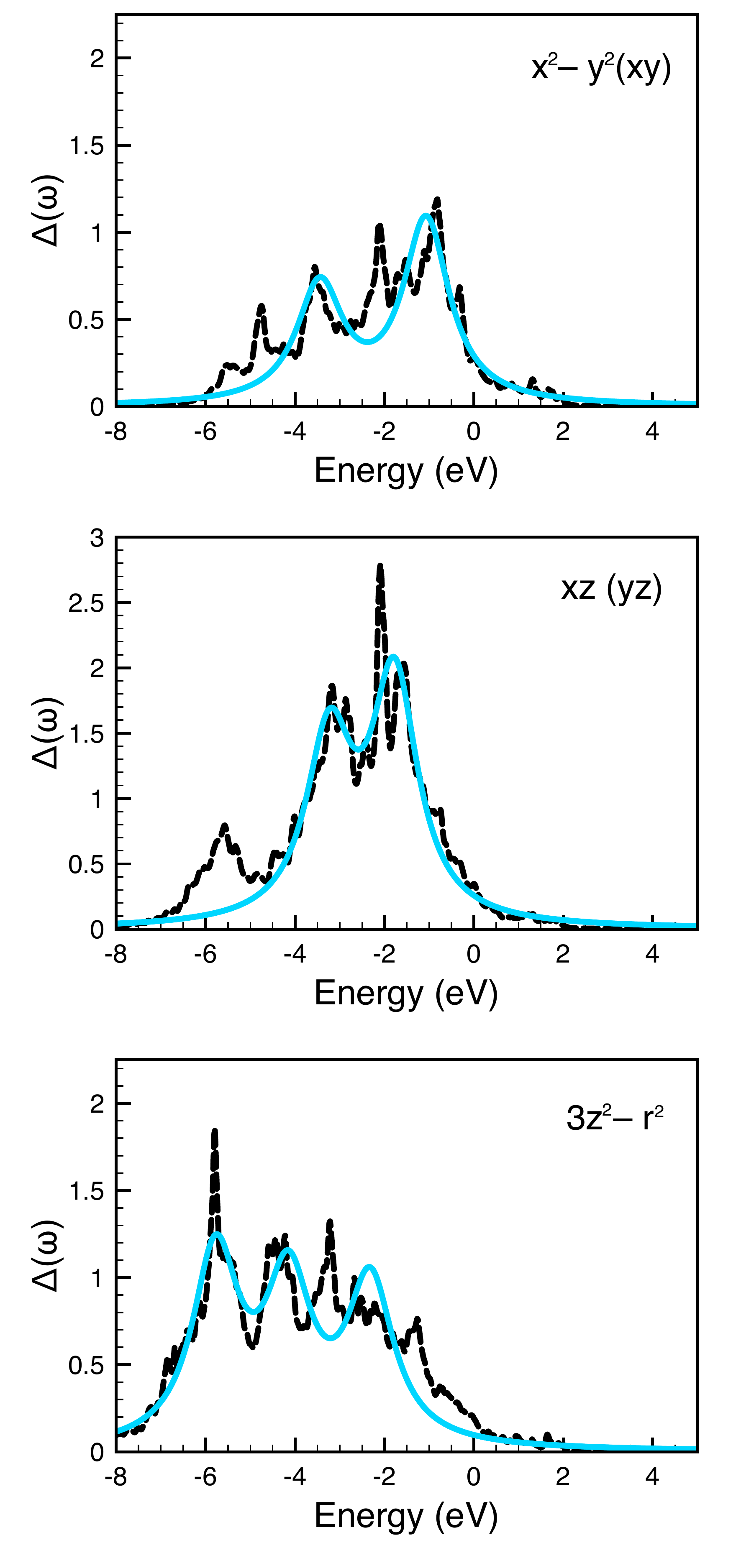}
 \caption{The imaginary part of the hybridization functions (dashed line) obtained from LDA calculations in comparison with fitting results by using Eq.
 \ref{eq:hybparameter} (blue solid lines).\label{hybminimization}}
\end{figure}

\noindent{\it Definition of the parameters.}  
The energies $\epsilon_p$ and hoppings
$V_{ip}$ were calculated within the minimization of the LDA hybridization functions presented in Fig.\ref{hybminimization} by using the following expression,
\begin{equation}
\Delta_{i}(\omega) = \sum^{N_{p}}_{p=1} \frac{|V_{ip}|^{2}}{\omega - \epsilon_p},
\label{eq:hybparameter}
\end{equation}
where $N_{p}$ is the number of the effective orbitals describing the surface.
The main limitation of the exact diagonalization approach is the number of the effective orbital in the electronic Hamiltonian.
In our study we have simulated five $3d$ orbitals of the adatom. Depending on the symmetry each impurity orbital is connected  with a certain number of the surface levels.
For instance, we use $N_{p} = 2$ for $xz (yz)$ and $x^2-y^2 (xy)$ orbitals.
 Since the hybridization function of $3z^2-r^2$ orbital demonstrates more complicated structure than others then for
this orbital we used $N_p = 3$ bath states. Thus the total number of the effective orbitals in Eq.(1) is equal to 16, which corresponds to the maximal occupation of 32 electrons.

The impurity orbital energies, $\epsilon_{i}$ were varied within the expression for the impurity bath Green's function
\begin{eqnarray}
G_{i} (\omega) = [\omega - \epsilon_{i} - \Delta_{i} (\omega)]^{-1}
\end{eqnarray}
to reproduce the LDA occupations for $3d$ states of iron described in the previous section. 

In turn, the elements of the Coulomb interaction matrix, $U_{ijkl}$ were defined by using effective Slater integrals that related to the averaged  on-site Coulomb interaction, $U_{d}$ and intra-atomic exchange interaction $J_{H}$ as described in Ref.\onlinecite{0953-8984-9-4-002}. 

\begin{figure}[!h]
\centering
   \includegraphics[width=0.3\textwidth]{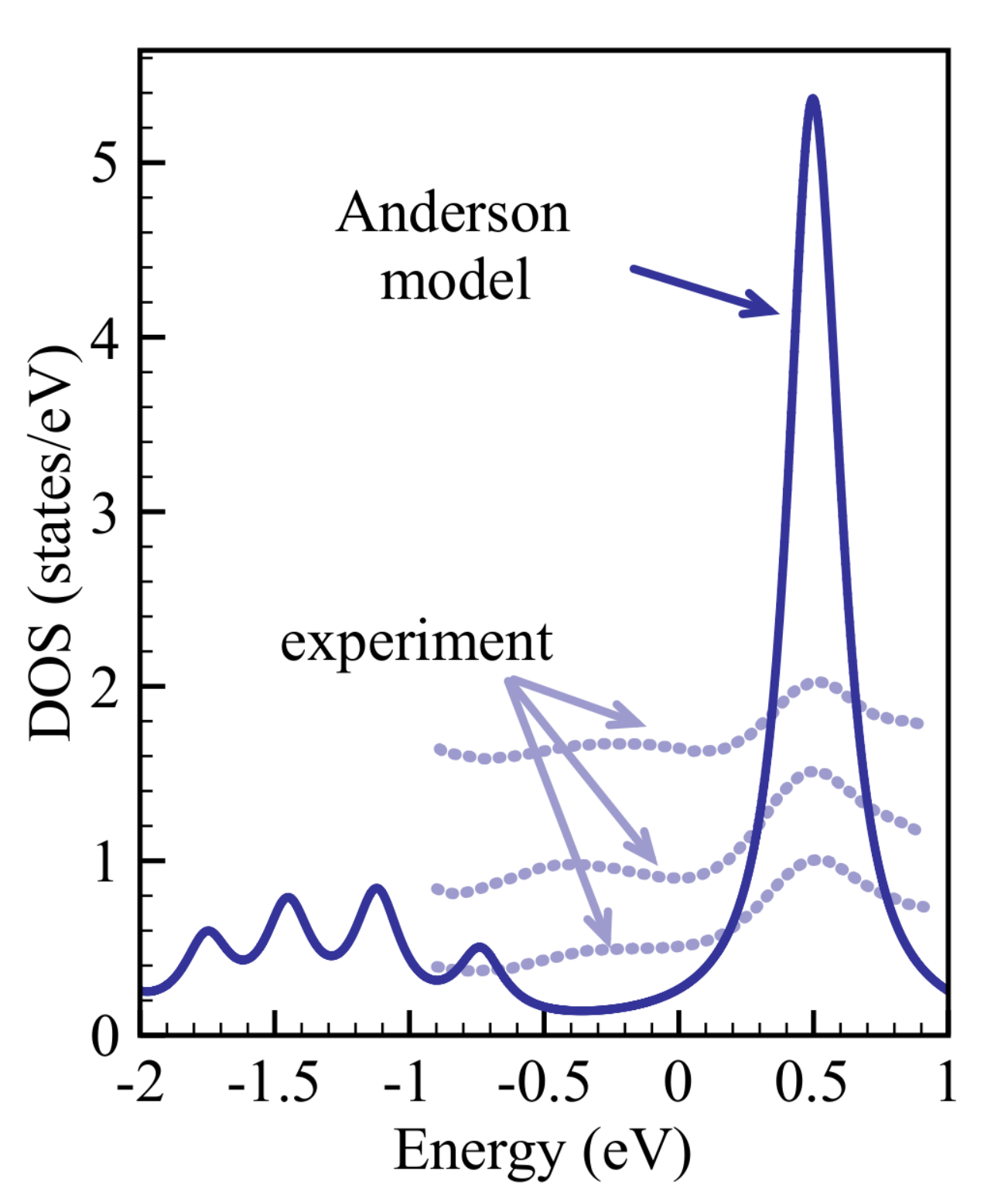}
 \caption{The spectral function of the iron impurity atom calculated by using the exact diagonalization approach (blue line). The dotted lines correspond to the densities of states determined in the STM experiment Ref.\onlinecite{Crommie} with different tips. \label{doscomparison}}
\end{figure}

\noindent{\it Correlated spectral functions.} 
The constructed Anderson model was solved by using finite-temperature exact diagonalization solver. To analyze the excitations near the Fermi level we have calculated 26 lowest eigenvalues and eigenvectors of the Anderson Hamiltonian. The main result of our investigation that is the comparison of the correlated spectral function and experimental densities of states extracted from the STM spectra \cite{Crommie} is presented in Fig.\ref{doscomparison}.   One can see that the position and width of the theoretical peak are in excellent agreement with the experimental results. The analysis of the partial densities of states shows that all the $3d$ orbitals equally contribute to the peak at 0.5 eV above the Fermi level. 

We found that the variation of $U_{d}$ from 2 eV to 4 eV and $J_{H}$ from 0.6 eV to 0.8 eV does not significantly change the correlated spectrum near the Fermi level.  One should mention that the choice of the chemical potential, $\mu$ is of crucial importance, since it controls the position of the resonance above the Fermi level. If one chooses the value of $\mu$ in such a way to reproduce the LDA total number of $3d$ electrons that is 6.5, then the resonance is at 0.3 eV. The solution with correct position of the resonance at 0.5 eV above the Fermi level corresponds to 6.13, which agrees with the LSDA solution.

\begin{figure}[t]
\centering
   \includegraphics[width=0.4\textwidth]{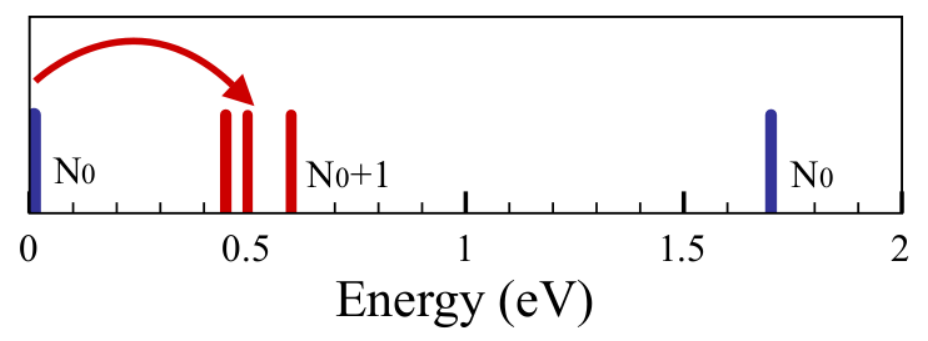}
 \caption{The calculated spectrum of the Anderson model. Arrow denotes the excitation corresponding to the peak at 0.5 eV above the Fermi level. N$_{0}$ is the number of the electrons in the cluster, N$_{0}$ = 27.\label{Andersonspectrum}}
\end{figure}

\begin{figure}[b]
\centering
   \includegraphics[width=0.42\textwidth]{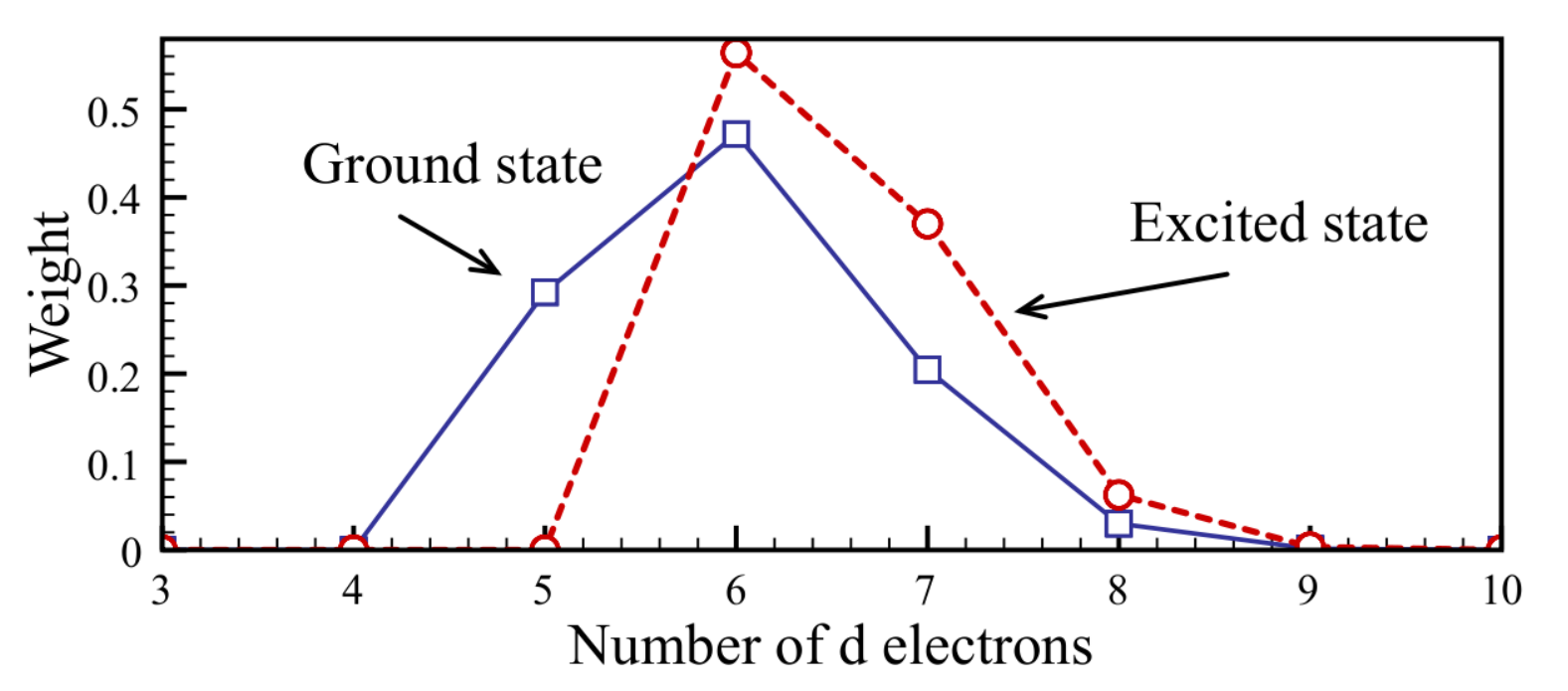}
 \caption{Weights of the impurity atomic configurations comprising the ground and first excited states.}
 \label{weight}
\end{figure} 

To understand the microscopic origin of the peak we analyzed the low-energy spectrum of the Anderson model presented in Fig.\ref{Andersonspectrum}. 
There are three excited states at 0.48 eV,  0.49 eV and  0.56 eV. The excitations from the ground state to first excited states correspond to the experimental resonance above the Fermi level. These excitations lead to the change of the total number of the electrons in the system that is increased by one.
The change of the impurity properties at this transition can be traced by calculating the distribution of the atomic configurations of the impurity for each eigenstate (Fig.\ref{weight}).  
For the ground state we obtain $0.3 d_{0}^5 + 0.47 d_{0}^6 + 0.2 d_{0}^7 + 0.02 d_{0}^8$, where the subscript index denotes the ground state. 
The iron impurity is mainly in $d^6$ configuration, which follows from the Hund's rules for isolated atom. A strong coupling with the substrate leads to the valence fluctuations\cite{Haule} to $d^5$ and $d^7$ configurations within the ground state.  In turns, the first excited state at 0.48 eV is characterized by the following composition of the atomic configurations $0.56 d_{1}^6 + 0.38 d_{1}^7 + 0.06 d_{1}^8$. Comparing the computed compositions for the ground and first excited states we observe the redistribution of the atomic configuration weights of the impurity, which corresponds to the $d_{0}^{5} \rightarrow d_{1}^{6}$ and $d_{0}^{6} \rightarrow d_{1}^{7}$ transitions.

Experimentally, there is also a shallow rise in the density of states at approximately -0.35 eV, which can not be reproduced by using the LSDA approach (Fig.\ref{LSDA}).  This experimental feature can be associated with the excitation at -0.7 eV in the theoretical spectrum of the Anderson model. The latter has the symmetry of the in-plane $xy$ and $x^2-y^2$ states.

Thus we obtain two different descriptions of the experimental STM resonance at 0.5 eV for Fe/Pt(111).  The first one is the LSDA solution where the peak is the result of the splitting between spin-up and spin-down $3d$ states of the adatom. On the other hand the solution of the five-orbital impurity Anderson model revealed a complex multiplet structure of the iron atom at low temperatures, which is the result of the interplay between intra-atomic exchange coupling and strong impurity-surface hybridization.   

Importantly, one can find a connection between the LSDA and Anderson model results. For an isolated atom with a partially filled shell it was shown by Slater \cite{Slater} that the difference between one-electron energies with spin-up and spin-down is proportional to the terms that remain even if there is no difference between spin-up and spin-down orbitals. Based on this result he found a connection between the spin-splitting energy computed in an one-particle approach and the difference between the average energy of all multiplets with  $\mathcal{S}$ and those with $\mathcal{S} -1$ (for the isolated iron $\mathcal{S}$=2) calculated by using a many-particle method. In our case these results should be revised by taking into account a strong hybridization with the surface, which leads to the valence fluctuation in the Fe/Pt(111) system. Since a further quantitative comparison of the LSDA and Anderson model energies requires the calculation of high-energy ($\sim$ 3-5 eV ) eigenstates of the Anderson Hamiltonian we left it for a future investigation.

\begin{figure}[!b]
\centering
   \includegraphics[width=0.42\textwidth]{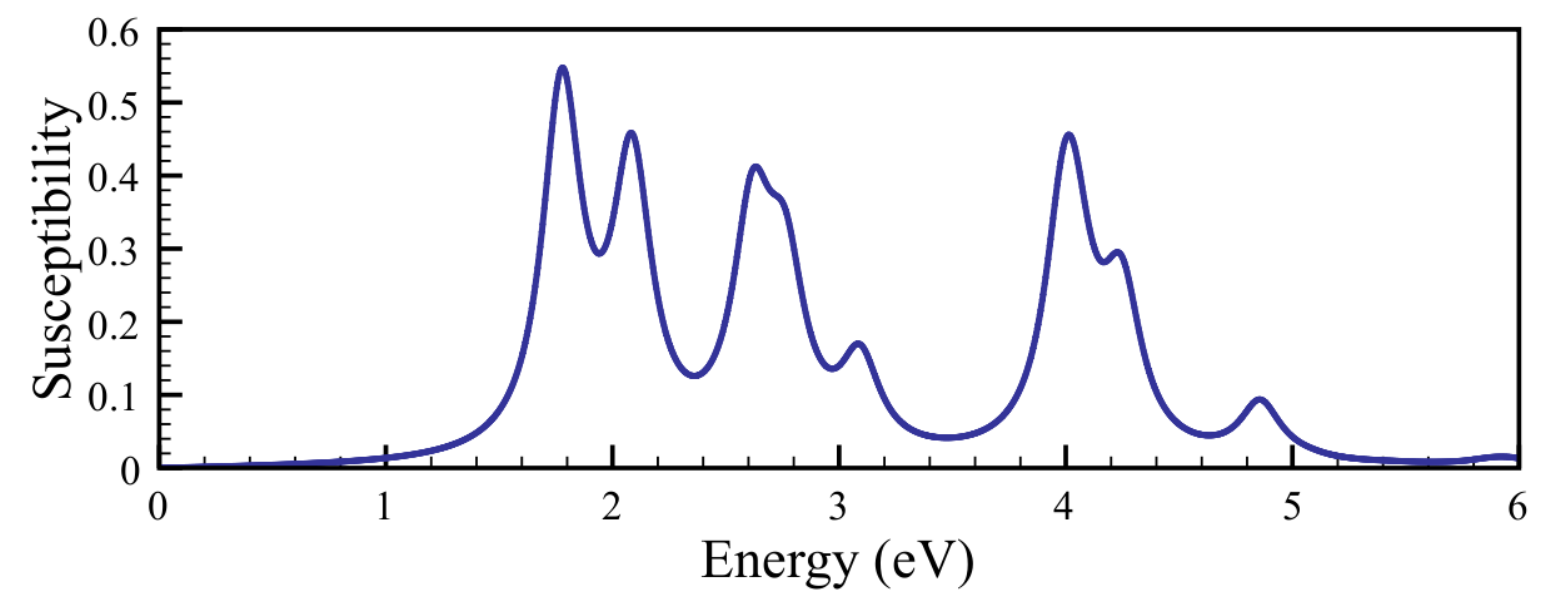}
 \caption{Spectral function of the total spin-spin susceptibility calculated by using Eq.4.}
\label{suscfig}
\end{figure}

\noindent{\it Magnetic susceptibility.} To complete the picture of the magnetic properties of Fe/Pt(111 ), we calculated the spin-spin susceptibilities for $3d$ states of the iron adatom,
\begin{equation}
\chi^{zz}_{i}(\omega) = \frac{1}{Z} \sum_{nn'} \frac{|\left<n'|S_{i}^{z}|n\right>|^{2}}{\omega+\imath\delta + E_n - E_{n'}}\left(e^{-\beta E_{n}} -
e^{-\beta E_{n'}}\right),
\label{susc}
\end{equation}
where $Z$ is the partial function, $E_n$ is the eigenvalue of the Hamiltonian, Eq.(1). One can see that the lowest resonance in the spectral function of $\chi(\omega)$ is at 1.7 eV (Fig.\ref{suscfig}). From Fig.\ref{Andersonspectrum} it follows that the corresponding excitation does not change the total number of the electrons in the system. On the level of the correlated density of states (Fig.\ref{doscomparison}) it refers to the excitation below the Fermi level at the corresponding energy. 

\section{Iron-hydrogen dimer}
\begin{figure}[!t]
\centering
   \includegraphics[width=0.42\textwidth]{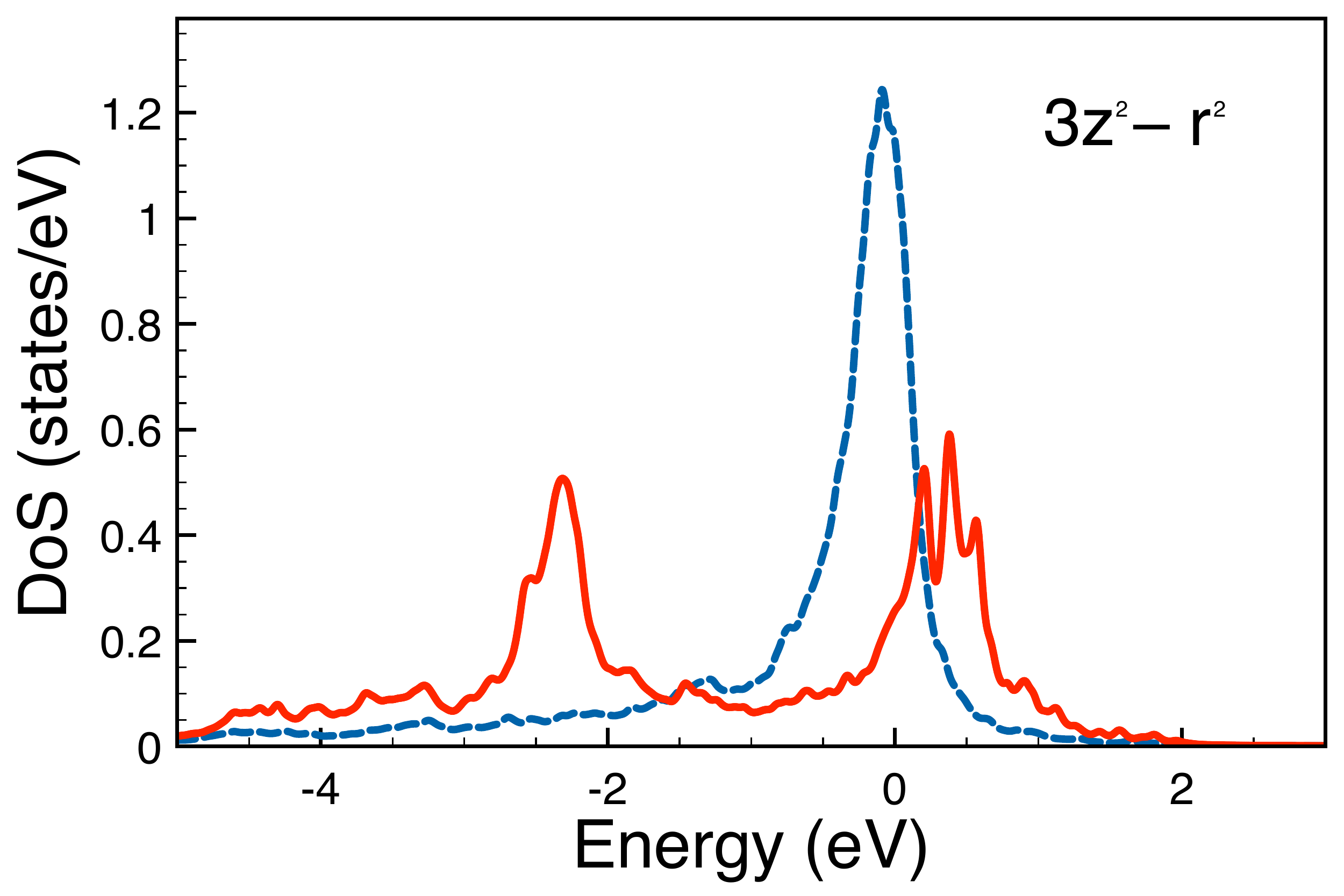}
 \caption{Comparison of the $3z^2-r^2$ densities of states calculated for Fe/Pt(111) (blue dashed line)  and FeH/Pt(111) (red solid line) by using the LDA method.}
 \label{FeHdoslda}
\end{figure}

In the previous works \cite{Bode, Wiesendanger2,Wiesendanger3} it was proposed that the resonance above the Fermi level can be used for detection of the iron atoms in surface nanostructures.
To check this proposition we consider the situation when the iron atom is hidden from the STM tip by another atom (Fig.\ref{structure}, right). Such a configuration can be realized experimentally.\cite{nature} For that the hydrogen atom was deposited atop the iron impurity. By using the first-principles molecular dynamics simulations the equilibrium distance between H and Fe atoms was obtained to be 1.6 \AA. 
Our LDA calculations show that the coupling with hydrogen does not change the iron partial densities of states for the $xy$, $x^2-y^2$, $xz$, $yz$ states. 
At the same time, the $3z^2-r^2$ atomic orbital of Fe and $s$ orbital of H form a molecular orbital.  
From Fig.\ref{FeHdoslda} one can see that the bonding and antibonding states are located at -2.4 eV and +0.5 eV, respectively. 

The Anderson model describing the FeH/Pt(111) is given by
\begin{equation}
\mathcal{H}_{FeH} = \mathcal{H}_{Fe} + \mathcal{H}_{H-Fe} + \mathcal{H}_{H},
\end{equation}
where 
$\mathcal{H}_{H} = \sum\limits_{s\sigma} \epsilon_{s} n_{s\sigma} $ describes $s$-states of hydrogen.
$\mathcal{H}_{H-Fe}$ corresponds to interactions between $s$-states of hydrogen and
$d$-states of iron and can be written as follows:
\begin{equation}
\mathcal{H}_{H-Fe} = \sum_{si\sigma} \left( V_{is}d^{\dagger}_{i\sigma}a_{s\sigma} + H.c.\right).
\end{equation}
Here $\epsilon_{s}$ is energy of hydrogen state, $V_{is}$ is the hopping between Fe and H states, $a_{s}^{\dagger}$ ($a_{s}$) is the creation
(annihilation) operator for H states.

To define the parameters of the Anderson model describing the FeH/Pt(111) system we assume that $s$-states H is hybridized only with $3z^2-r^2$ states of iron atom.
Based on these assumption we minimize the hybridization functions of the $3z^2-r^2$ orbital with two bath orbitals for Pt and single effective orbital for
hydrogen. The hopping integral between hydrogen and iron orbitals equals to $V_{is}$ = 1.56 eV.

\begin{figure}[t]
\centering
   \includegraphics[width=0.42\textwidth]{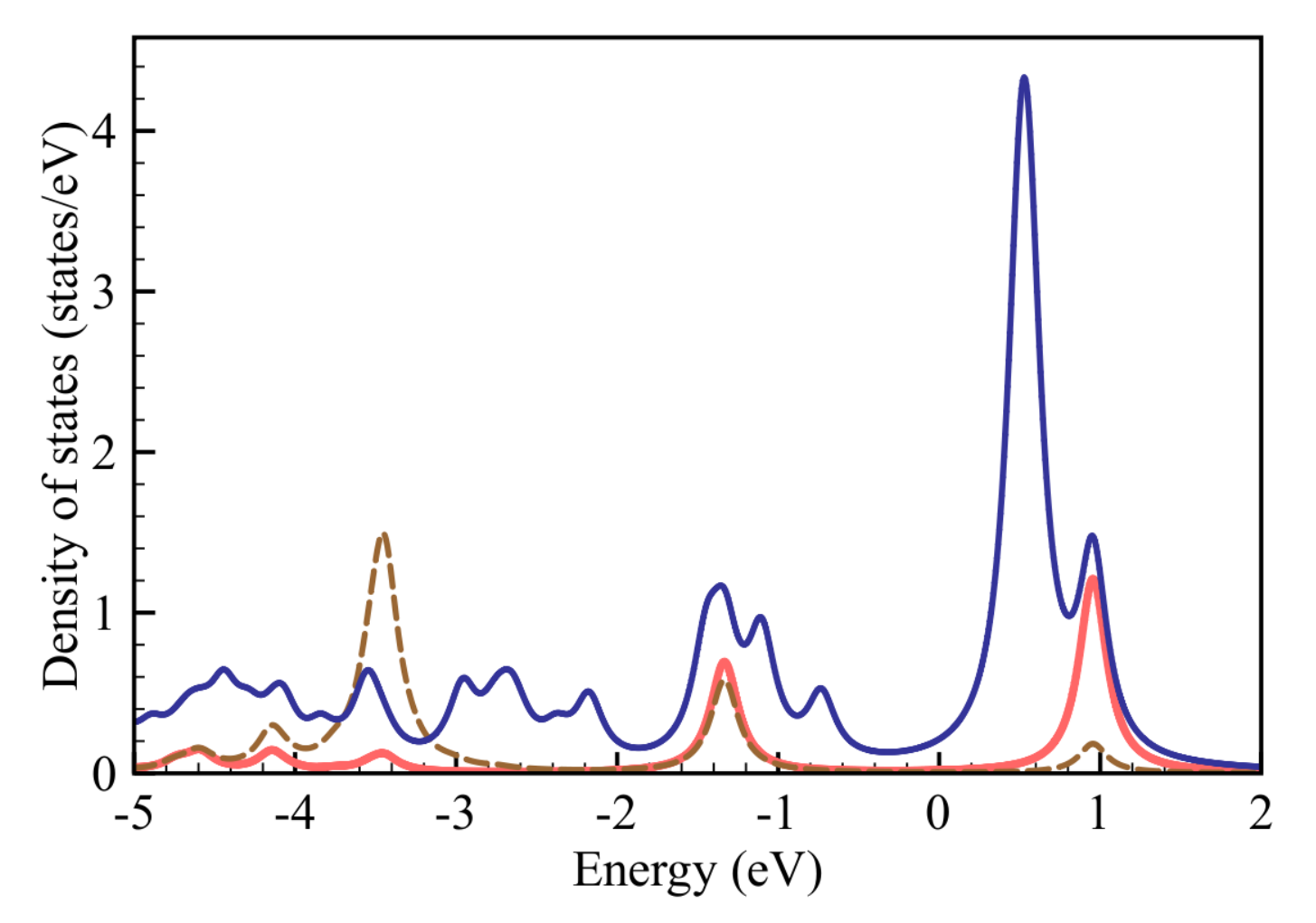}
 \caption{Correlated densities of states for FeH/Pt(111). Blue line corresponds to the summary density of $xy$, $yz$, $xz$ and $x^2-y^2$ states. Red line denotes $3z^2-r^2$ states. Dashed brown line denotes the spectral function of the hydrogen atom deposited atop the iron impurity.}
 \label{FeHcorrdos}
\end{figure}

The correlated spectral functions of the FeH/Pt(111) system is presented in Fig.\ref{FeHcorrdos}. One can see that there two resonances above the Fermi level. Similar to Fe/Pt(111) the first one is at 0.5 eV which is due to the $xy$, $x^2-y^2$, $yz$ and $xz$ states. The second one at 1 eV is originated from the iron-hydrogen hybridization. Thus, in contrast to Fe/Pt(111), the resonance of the $3z^2-r^2$ states is shifted to +1 eV. We observe the  excitation at the same energy in the hydrogen spectral function that will mainly contribute to the STM spectrum in the simulated configuration.    

\section{Conclusions}
In summary, we have studied the electronic structure and magnetic properties of the Fe/Pt(111) system by means of the first-principles calculations based on the density functional theory and model Anderson impurity Hamiltonian that was solved by using exact diagonalization approach. It was found that both approaches successfully reproduce the main experimental feature that is the peak at 0.5 eV above the Fermi level in the STM spectrum. While within LSDA the peak originated from the spin splitting of the $3d$ shell, on the level of the Anderson model the excitation is associated with the valence fluctuations between atomic configurations of iron. The connection between these results on the basis of the seminal work by Slater \cite{Slater} is discussed.  In addition, we predict the shift of the STM resonance to 1 eV in FeH/Pt(111).

\section{Acknowledgments}
The hospitality of the Institute of Theoretical Physics of Hamburg University is gratefully acknowledged. 
The work is supported by the grant program of the Russian Science Foundation 14-12-00306.

\end{document}